# Field Tests of High-Density Oil-Based Drilling Fluid Application in Horizontal Segment

Youcheng Liu[1]

*Technology Department, Great Wall Drilling Company, CNPC, Beijing 100101, China*

**Abstract**

The operation experience of horizontal segment well in the southern Sichuan area within the hard brittle shale layer of Longmaxi formation demonstrates that the high-density oil-based drilling fluid (HDOF) induced down-hole working condition, which is typically high-temperature, high-pressure and solid-phase-dominant, will directly impact the stability of the wellbore wall. Focused on the impact of bottom-hole osmotic pressure and solid phase content and a series of density-reduction field tests, we found some helpful countermeasures. In this paper, we study and discuss the impact of high osmotic pressure and operation measures, and provide details of the field tests which indicate that maintaining a higher osmotic pressure at the bottom-hole can reduce the formation pore pressure of the bottom hole, widen the fluid density window, and maintain the stability of the wellbore wall. By improving the HDOF formula, such as using new treatment agents (e.g. polymer modified nano-sealing agent, ultra-micro barite weight additive) and optimizing the solid particle size distribution of HDOF, we can effectively address the down-hole problems. The field test of "pressure-controlled drilling + density reduction" in horizontal drilling operations is therefore proved to be capable of reducing the complexity of down-hole faults, improving the rate of penetration, and saving the overall cost.

**Keywords: field test; high-density; osmotic pressure; solid phase sag; horizontal segment**

---

[1] Corresponding author

E-mail address: lyc7193@163.com

# 1. Introduction

The application of water-in-oil (W/O) drilling fluid in the horizontal drilling process has been gradually becoming the mainstream technology. High-density oil-based drilling fluid (HDOF) enables the pressure balance between deep formation and bottom-hole, mitigating the risk of well-control and possessing the significant capacity of rock carrying and wellbore wall stabilizing. HDOF is typically applied to prevent the sloughing or shrinkage of the wellbore. In general, the stability of the wellbore wall in shale formation depends on the mechanical and chemical coupling effects, and the appropriate increase in the density of drilling fluid can often effectively prohibit the collapse of the wellbore wall. The Longmaxi formation, one of the most typical target layers of Southern Sichuan area shale gas, is dominated by high-silicon and hard-brittle mud shale (Argillutite) [1, 2, 3, 4, 5], and the static pressure of high liquid column formed by HDOF will induce or aggravate the micro-fracture development surrounding the hard-brittle mud shale wellbore wall. The pressure fluctuation caused by the tripping work and the starting/stop pumping may result in the repeated opening and closure of micro-fractures, which further induces the stress concentration and stress releasing along the crack surface. The shear displacement of the extended crack surface occurs, which leads to large variety of complex down-hole circumstance [6, 7, 8].

To date, the operation barriers introduced by the recent drilling difficulties in Southern Sichuan area show some common features related to high-density drilling fluid. Complex situations such as sloughing and redressing will be significantly aggravated as the drilling fluid density increases. Although increasing the drilling fluid's density is capable of maintaining short-term stability of the wellbore wall, the subsequent filtrate intrusion will eventually reduce wellbore wall's stability. Therefore, a positive feedback loop of "wellbore wall instability - increasing density - short-term stability wall - aggravating the filtrate percolation - collapse deterioration" would severely impact the drilling operations. The higher the fluid density going up, the worse the wellbore wall stability circumstance going worse, and the more severe the wellbore wall sloughing.

The operation experience of horizontal well section in the southern Sichuan area within the hard brittle shale layer of Longmaxi formation demonstrates that the HDOF-based down-hole working conditions, which is typically with high-temperature, high-pressure and solid-phase-dominant, will directly and negatively impact the stability of the wellbore wall and specifically increase the risk of sloughing and circulation loss. For the drilling fluid system with barite as the weighting additive, the volume fraction of solid phase content can reach 45% or higher when the density is above 2.3 g/cm3 [9, 10, 11], which gives rise to the difficulty of controlling the rheological parameters, stabilizing the wellbore wall and maintaining the rate of penetration (ROP).

The emphasis of this paper focuses on high density W/O drilling fluids treated with barite weighting additive and with a density range of 1.8 g/cm3 to 2.3 g/cm3 [10, 11, 12, 13]. As a field test case, we focus on the horizontal wellbores of Longmaxi formation, Weiyuan block in Southern Sichuan area, aiming to address the HDOF implementation strategy and study the aspects of osmotic pressure and high solid phase content performance in horizontal segment. We primitively verify the conclusion through the field test of the density reduction drilling.

# 2. Impact of high osmotic pressure

To date, the research related to osmotic pressure becomes mature in the mechanics-chemistry coupling field regarding the wellbore wall stability. Due to the different activity of water content in the formation and drilling fluid, the phenomenon of fluid entering and leaving the formation in the drilling process, would affect the stability of the wellbore wall. The method of reducing the hydration of mud shale formations was firstly applied to oil-based drilling fluid system and achieved remarkable results by mitigating water activity of drilling fluid. Considering the bottom-hole pressure profile condition prescribed by the safety equivalent-density window (we only examined the equivalent-density window between the pore pressure and the fracture pressure), the static bottom-hole pressure relationship in the elastic porous formation can be briefly described as Eq. 1:

$$P_k + P_\pi \leq P_w \leq P_f \quad (1)$$

Where, $P_k$ is the formation's pore pressure, $P_\pi$ is the osmotic pressure, $P_w$ is the bottom-hole pressure (which is balanced with the pressure of the bottom-hole fluid column pressure), and $P_f$ is the fracture pressure. According to the research of [14], the coupling value $P_k + P_\pi$, known as "equivalent pore pressure", can also be considered as the minimum fluid column pressure.

The pressure relationship presented in Eq. 1 can be characterized as the corresponding equivalent density relationship shown in Eq. 2:

$$\rho_k + \rho_\pi \leq \rho_w \leq \rho_f \quad (2)$$

Where, $\rho_k + \rho_\pi$ is the equivalent density of $P_k + P_\pi$, $\rho_w$ is the equivalent density of $P_w$, and $\rho_f$ is the equivalent density of $P_f$.

Mud shale pressure transferring experiments show that as $P_\pi$ changes, it will change both magnitude and distribution of $P_k$ on the borehole [15], and the equivalent pore pressure $P_k + P_\pi$ will also change accordingly. According to Eq. 1 and Eq. 2, the decreasing of $P_\pi$ will result in decreasing of $P_k + P_\pi$, and $\rho_k + \rho_\pi$ will decrease subsequently. At this point, the lower limit of the $\rho_w$ goes down and the room of density-reduction can be expanded. According to the Gibbs–Donnan equilibrium theory [16], water phase osmotic-movement driving force can be generated due to the concentration difference at the semi-permeable membrane interface between solutions locating in both sides. By changing the water activity in the oil-based drilling fluid system and changing the direction of water transfer, the stabilization of the wellbore wall can be achieved in the drilling operation. We represent the relationship between osmotic pressure and water as Eq. 3:

$$P_\pi = -\delta_m \frac{RT}{V_m} ln \frac{A_{shale}}{A_{mud}} \qquad (3)$$

Where, $P_\pi$ is osmotic pressure, $\delta_m$ is the efficiency of semi-permeable membrane, $R$ is the gas constant, $V_m$ is the molar volume of water, $A_{shale}$ is the shale water phase activity, and $A_{mud}$ is the water activity of the drilling fluid. As shown in Eq. 3, when $A_{shale} > A_{mud}$, the osmotic pressure $P_\pi$ turns to be negative, at which point water penetrates through the rock formation into the wellbore and stabilizes the wellbore wall. When $P_\pi$ becomes negative, the decrease of $P_\pi$ gives rise to increasing of the absolute value of the osmosis pressure. Therefore, at the range where $P_\pi$ is negative, maintaining a high osmotic pressure when horizontal drilling can not only effectively prevent shale hydration expansion of wellbore wall, but also contributes to the reduction of the drilling fluid density.

By assuming that the formation's water phase activity $A_{shale}$ and the semi-permeable diaphragm efficiency $\delta_m$ as 1 and 1.0, respectively, the bottom-hole osmotic pressure can reach 69 MPa when the mass fraction of ($CaCl_2$) in the oil-based drilling fluid water phase reaches 31; when the mass fraction of $CaCl_2$ in the water phase reaches 40%, the bottom-hole osmotic pressure can reach 111 MPa, which will be sufficient for dehydrating contraction and peeling of shale formations surrounding the wellbore [17]. In common cases, when using W/O drilling fluids, the semi-permeable diaphragm efficiency and the osmotic pressure produced by the difference in water phase activity between both sides of wellbore wall are relatively high. It is therefore important to monitor the change of drilling fluid oil-to-water ratio in the operation procedure and timely adjust the activity of water phase, which can accordingly prevent the water-loss of wellbore wall.

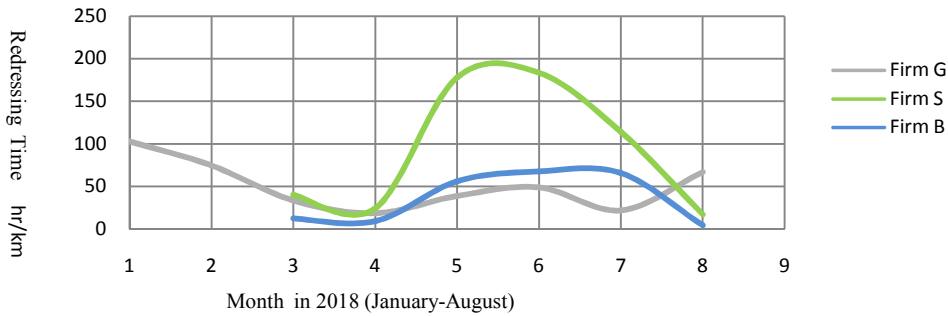

**Figure 1** Redressing time of partial horizontal drilling operation in Weiyuan block

It's worth noting that at the early stage of Weiyuan block development (before 2019), W/O fluid used the CaCl2 solution with the concentration of 18%-25% as water phase and controlled the water activity range at 0.82-0.72, which resulted in relatively high down-hole failure rate, high frequently of circulation loss and long redressing time. Figure 1 shows the stats of the redressing time (hours per kilometer) of some drilling crews obtained back in 2018. After increasing the concentration of CaCl2 in water phase to 35%-45%, the activity range is controlled in the range of 0.3-0.5, which significantly reduced the down-hole failures such as circulation loss and redressing time. In addition, appropriate decreasing the drilling fluid density can also improve the quality of drilling fluid tremendously.

The earlier stage operation experience indicates that water phase activity equilibrium in high-density drilling operation enables high bottom-hole osmotic pressure, which reduces of the bottom-hole pore pressure, widens the equivalent density window (Eq. 2) and maintains of the wellbore wall stability. Increasing the water activity in W/O drilling fluid and the bottom-hole osmotic pressure is conducive to increasing the room of density-reduction. When making an operation scheme, engineers should sufficiently consider the influence of water activity in operation with HDOF by keeping a suitable activity level in water phase, determining the fluid preparation concentration of $CaCl_2$, and maintaining the favorable conditions of high-level bottom-hole osmotic pressure.

The aforementioned countermeasures to increase the bottom-hole osmotic pressure are the result of optimization with a given design density. The safe density window design for drilling operation scheme depends on the formation collapse pressure, pore pressure, leakage pressure, fracture pressure and computational model[18]. Nevertheless, the design of drilling fluid density in Weiyuan area is by far based on the experience of adjoining wells, and the method of providing above parameters has not been accurately validated. In the future, it is necessary to further conduct pressure profile analysis in combination with the geological features and actual drilling experience of the specific blocks, to deepen the understanding of the stratigraphic fissure distribution features and its changing, and to conduct the performance optimization of HDOF according to the new circumstance of density design, so that we can mitigates this difficulty confronting the shale gas exploration.

## 3. Effects of high solid phase content
### 3.1 Solid phase sag

The sag of high-density drilling fluid at the horizontal segment of the well has its own particularity and imperceptible concealment [19]. Due to the lack of balance of vertical shear, the probability of gravity sedimentation at the horizontal segment is much higher than that of vertical segments. The flow-rate fluctuation of high solid phase content drilling fluid in the annular space of horizontal segment would induce the solid phase movement difficulty, and the probability sag of the barite and cuttings will subsequently increase. When the annular volume flow rate drops down and the Reynolds number of drilling fluid flows decreases, the flow-pattern converts to advection or flat flow and the sag will be significantly intensified, thus the cuttings carrying process of turbulence with high pump pressure in horizontal drilling operation is beneficial for mitigating the solid phase sag. The target layer in Weiyuan area is mostly belongs to the Longmaxi formation, which belongs to the hard-brittle shale high-pressure gas reservoir and is accompanied by micro-fracture development [20]. Since the basic research work of the formation core in Weiyuan area has not been well-studied, the distribution and quantity of the original micro-fractures in rock are not comprehensive, and the targeted lab-research is difficult to conduct.

The ultra-high static fluid pressure formed by HDOF directly leads to faster liquid immersion into the borehole surrounding rock. On one hand, The drilling fluid filtrate may flow into pores of the rock and micro-fractures, which accelerates the cracking and peeling of the mud shale along the direction with lower shearing stiffness and reduces the strength and hardness of the rock around the borehole over time. Eventually, it leads to the rupture of the wellbore wall. On the other hand, the enhancement of the pressure filtration effect leads to the increasing of the solid phase volumetric content in drilling fluid and negatively impacts the drilling fluid's rheological properties. The significant mechanical grinding effect between barite particles and drilling cuttings or bentonite can lead to a great increase in the accumulation of harmful low-density solid phases such as clay or drill chips than conventional drilling fluids; at the same time, the barite particles grain size becomes thin and the loss of them will subsequently increase.

It is difficult to maintain the dispersion balance in the environment with solid phases such as weighting agent and cuttings, which can induce the overall dynamic sag (solid phase) in horizontal segment at any time and even drill pipe sticking. Partial degradation of additive molecules such as emulsifier, cuttings-wetting agent in the high temperature and high pressure (HTHP) at bottom hole not only leads to the sag of weighting agent and drilling cuttings, but also weakens the drilling fluid rheology performance, for example, the fluid viscosity leap because of the solid phase particle-aggregation. High-temperature suitable emulsifier and wetting agent should be selected to ensure less boundary tension or smaller contact angle with a stable transport carry function, so as to strengthen the wellbore-cleaning function of drilling fluid. By optimizing the particle size and particle distribution of oil-based drilling fluid, the effects of solid gravity sedimentation can be mitigated. The follow-up operation should promote the field application of wetting agent dispersant and provide support for efficient removal of harmful low-density solid phase and the stable and controllable oil-based drilling fluid technology.

As a well-bore cleaning countermeasure, the horizontal segment drilling fluid flow-rate is usually controlled at 30 - 50 liter/second or higher. Within the flow-rate range, the drilling fluid along the borehole performs the flow-pattern of turbulence. This countermeasure can achieve the goal of cuttings conveying. But meanwhile, the formation and quality of mud cake will also be influenced by high fraction of solid phase such as drilling cuttings and weighting agent, filtrate loss (FL) control agent and so on. The sustained erosion of high flow-rate to the wellbore wall and abnormal increase of dynamic FL may negatively affect the formation and quality of mud cake as well. The research of hard-brittle shale formations in North America region through the field examples, lab-experiments and numerical simulation results by P. J. Mclellan et al [21] shows that for similar fracture formations, the drilling fluid cannot even successfully form an effective mud cake in conditions of high density, high Reynolds number, high solid phase coupling with poor solid-phase particle size distribution.

### 3.2 Solid phase particle size optimization

The thinness, brittleness and fissure development of the surrounding rock-layer are the internal causes of the collapse of the well leakage. Figure 2 Based on the fissure-measurement of the W**7-2 well core. The scanning electron microscope (SEM) observation surface is the unbroken longitudinal profile of Longmaxi formation, with a total of 523 fissure measurements, with holes and fissures sizes ranging from 104nm-244μm; the distribution of fissure: D10=0.70μm, D50=2.29μm, D90=8.91μm. According these data, it's supposed to design and optimize the drilling fluid solid phase Particle Size Distribution (PSD) based on "2/3 Bridging Theory". Table 1 shows the optimized distribution of HDOF granularity.

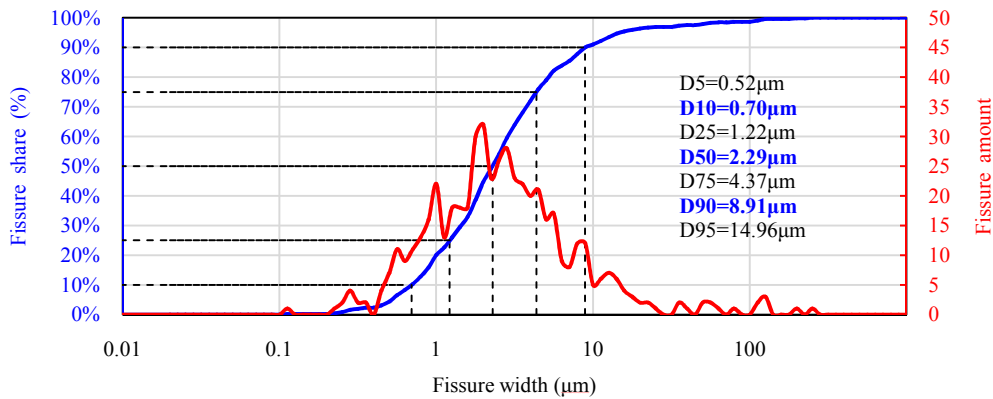

**Figure 2** The SEM Fissures analysis of Longmaxi formation: based on the core of well W**7-2.

The optimization of solid-phase particle size mainly includes the optimization of particle size and particle size distribution. The permeable formation wellbore can be divided into mud cake layer, bridge layer and penetration layer according to different depths, and the scale of micro-fractures is similar to the pores size of the bridge layer and penetration layer [17]. Under the effect of penetration pressure difference, the sealing agent particles will fill into the bridge layer, the penetration layer and the micro-fracture to achieve the function of sealing.

If the sealing performance of HDOF is insufficient, the sloughing-blocks would be difficult to control, which possibly leads to hard drilling trips. By optimizing the solid-phase particle distribution based on sealing agent and increasing the specific surface area of the solid-phase additives, the hazardous down-hole incident and damage related to the gravity sedimentation can be reduced. The particle size distribution of drilling fluid can be optimized by using the new-type nano-sealing agent. The quality of mud cake and the wellbore wall stabilization performance of drilling fluid can be improved and strengthened by choosing particle size with wide range distribution and promoting micro-nano scale sealing performance, respectively.

Table 1 Particle size distribution (PSD) optimization of high-density drilling fluid
(The optimization fluid = original drilling fluid + [1.5%FA-M] + [1.2%MN-2] + [1.2%RB-N] )

| Particle size parameters | Longmaxi seam crack hole size distribution | Particle size distribution of idealizing solid-phase | Particle size distribution of the original drilling fluid | Sealing agent particle size optimization scheme: concentration & type | | | Particle size distribution of the drilling fluid after optimization |
|---|---|---|---|---|---|---|---|
| | | | | 1.5% FA-M | 1.2% MN-2 | 1.2% RB-N | |
| D10 (μm) | 0.70 | 0.47 | 3.6 | 0.82 | 0.43 | 0.01 | 3.5 |
| D50 (μm) | 2.29 | 1.53 | 10.8 | 1.98 | 1.85 | 0.10 | 6.4 |
| D90 (μm) | 8.91 | 5.97 | 35.3 | 3.97 | 4.67 | 0.15 | 13.7 |
| Notes | Min: 0.104μm Max: 244μm | Based on the " 2/3 Bridging Theory" | Lack of corresponding sealing particles | Complement the flexible nano-micron sealing particles | Complement the rigid nano-micron sealing particles | Complement the nano sealing particles | Increased matching with the seam crack hole size |

In order to optimize the sag stability of HDOF, some operations have adopted ultra-micro barite additive with the particle median size $D_{50}$ at the mesh number of 6000 or more. This type of barite additive which has a smaller particle size and higher specific surface area is closer to the effect of colloidal dispersion system, so the suspension performance is good and the horizontal segment dynamic sag can be significantly mitigated. The technology has been widely applied in the operation of extended reach drilling and horizontal drilling segment of Schlumberger overseas projects, and has gradually become widely acknowledged [7]. Compared to conventional HDOF, the optimized HDOF system WARP of M-I company play better performance by reducing the viscosity approximately 10% and reducing the equivalent cycle density (ECD) approximately 50% in the Statfjord oilfield, North Sea area, Norway [22]. The rheological performance data after the use of ultra-micro barite (2 tons) in the W**54-3 well is shown in Table 2. The micro-particle of barite weighting additive has irregular geometry and the comparable size to the wellbore rock's micro-fissure which enables the functionality of bridge-sealing, strengthens the sealing performance of drilling fluid, reduces the erosion of the wellbore wall, and benefits the formation of the effective filter cake. Considering the solid-control loss and operation cost, operations practically adopt the way of combining the ordinary barite and the micro-powder barite, appropriately improve the mesh number of the desander and the vibrating screen at least 200. The operations should intermittently use centrifuges to keep the solid-phase content is within the scope of the control index; otherwise the rig should be required the barite powder recoverer (currently the standard equipped with ZJ50 or above drilling rigs). The amount of micro-powder barite agent filling can be controlled at about 30%-60% of the total mass amount of weighting agents. Accompanied with the application of the micro-powder barite and the nano-sealing agent, the additive-amount of FL control agent or viscosity enhancer in the fluid is supposed to be appropriately reduced. This countermeasure can help us maintain the overall solid phase content of the system, and meanwhile we can prevent the apparent viscosity (AV) leap, and stabilize the drilling fluid rheological index such as plastic viscosity (PV) and yield point (YP).

Since the first sub-layer of Longmaxi formation belongs to the breaking brittle layer, the sealing of the brittle formation is the key to solve the wellbore sloughing-blocks in this zone. By recently study, we proposed the scheme of multi-stage combination of rigid and flexible nano-micron sealing agent to solve the technical problem of micro-fracture developing in shale formation of Weiyuan block where traditional sealing agents failing to function effectively. Combined with the features of hard-brittleness and easy development of micro-fracture in mud shale wellbore wall of Longmaxi formations, the drill crew optimized the formulation of oil-based drilling fluids and increased the amount of high-quality nano-sealing agent. The new sealing scheme significantly improves the adaptability of HDOF in southern Sichuan.

Table 2 W**54-3 well oil-based drilling fluid performance indicators (after particle size optimization)

| | General specification | | | | | Rheological parameters | | | |
|---|---|---|---|---|---|---|---|---|---|
| Well depth | Density | Viscosity(M. Funnel) | FL(HTHP) | Mud cake | Solid content | PV | YP | YP/PV | Φ6 |
| m | g/cm$^3$ | s | ml | mm | percent | mPa·s | Pa | N/A | N/A |
| 2586 | 2.00 | 90 | 3.2 | 1.2 | 39 | 51 | 11.5 | 0.23 | 8 |
| 2804 | 2.01 | 99 | 2 | 1 | 39 | 52 | 15 | 0.29 | 10 |
| 3164 | 2.00 | 103 | 2.6 | 1.5 | 39 | 52 | 14.5 | 0.28 | 9 |
| Design specification | 2.00~2.10 | 90~110 | ≤3 | ≤2 | ≤40 | 50~65 | ≥11 | ≥0.18 | ≥8 |

The aforementioned technology is applied to 15 wells in Weiyuan block, which effectively improved the stability and reduced the sloughing blocks of the wellbore walls. At this point, the conventional wells (which is different from pressure-controlled drilling wells) density range is reduced from 2.05-2.20 g/cm3 to 1.95-2.02 g/cm3, and the redressing time is reduced from 104.4 hr/km to 29.62 hr/km, providing a guarantee of efficient shale gas field development. One of the example operation is W**55-3 well, whose actual drilling depth is 4668 meters and the horizontal segment is 1506 meters. After using the optimized HDOF in the third spudding operation, the well's horizontal segment drilling time has been effectively reduced to 17 days; throughout the entire operation process, trip-in and trip-out was smooth and the problems of redressing, sticking, serious sloughing-blocks and any other complex down-hole failure disappeared, which has greatly improved the time efficiency of operation procedures.

## 4. Density-reduction tests

As mentioned above, the safety density window design for drilling operation scheme depends on the formation collapse pressure, pore pressure, leakage pressure, fracture pressure and the corresponding computational model. The current design of drilling fluid density in Weiyuan blocks is largely based on the experience of adjoining wells, and the method of obtaining above pressure parameters has not been accurately confirmed. Therefore, we require a methodology that, on the basis of a given design density, the reasonable drilling fluid operation density should be moderately adjusted, explored and optimized, with a view to further enriching and accumulating the experience and data from the actual drilling operation. In order to comprehensively validate the impact of oil-based drilling fluid density and the effect of optimized water phase activity in oil-based drilling fluid and optimized solid phase distribution, the firm GWDC launched the field test of "pressure-controlled drilling + density reduction". The two platforms of W**33 and W**2 were equipped with pressure-controlled drilling equipment. The W**33-4 well and the W**2-8 well were taken the density reduction test after the procedure of well circulation loss failure, while the W**2-7 well started the whole-procedure implementation of the density reduction test at the beginning of the third spudding operation.

The first trip-in according to designed density of 2.0 g/cm$^3$ of the W**33-4 well was drilled from the depth of 2349 m. The sand return appeared normal, and there was no slough or circulation loss during the procedure until the circulation loss emerged at the depth of 2886.85 m. When drilling to the depth of 3314 m, (Landing point depth 3280 m) the well circulation loss emerged again and was accompanied with recirculation rate reduction and even drill string overflow while the pump shut down. When implementing the procedure of leakage plugging under pressure (while drilling), the re-loss of circulation was severe, and the concentration of plugging agent was up to 27%.

The density of fluid during leakage sealing decreased from 2.00 g/cm$^3$ to 1.92 g/cm$^3$. However, due to the pressure of the liquid column was relatively high, the leakage plugging measures failed to function effectively. Subsequently, we installed the pressure control facilities to implement the countermeasure (field test) of "pressure-controlled drilling + density reduction", reducing the fluid density to 1.82 g/cm$^3$ and keeping the sealing agent concentration at 2.5%. The leak was successfully plugged, and there was almost no sloughing blocks found on the vibrating screen surface during the pressure-controlled drilling procedure. The drilling operation was terminated after the drilling depth steadily reached 4680 m (the bore-hole segment length is 2331 m, and the horizontal segment length 1400 m).

In the subsequent operations, a total of 2 field test wells, namely the W**2-8 and W**2-7 wells, were completed. Geological data shows that the drilling tracks of these two wells' designed horizontal segment would pass through the broken formation zone, which may escalate the risk of circulation loss. Meanwhile, the horizontal drillings passed through the old wells, which kept very close distance even less than 300 meters: W**2-8 passed through between W**2-2 and W**2-3, and W**2-7 passed through between W**2-2 and W**2-1. The long-term exploitation of the old wells resulted in dropping of the pressure of the original formation. Eventually, this group of field tests obtained the anticipated results and achieved the desired objectives.

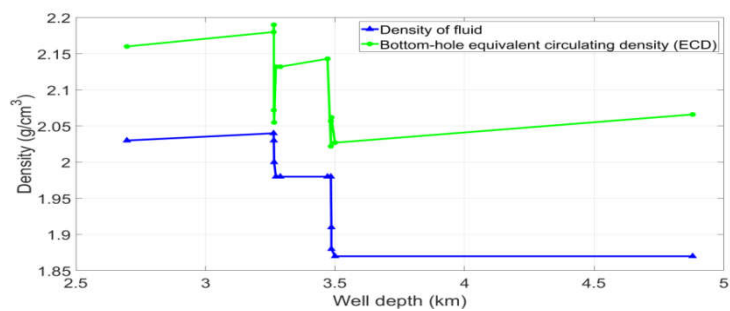

**Figure 3** The density curves of W**2-8 well (test well)

The drilling fluid density started with 2.03 g/cm$^3$ in third spudding of the W**2-8 well. A circulation loss emerged when drilling to the depth of 3267 m. After reducing the fluid density to 1.98 g/cm$^3$ and implementing the leakage plugging measure, the regular drilling was recovered. The circulation loss emerged again when drilling to the depth of 3484 m, with the constant drilling fluid density of 1.98 g/cm$^3$. The drilling fluid density was firstly reduced to 1.91 g/cm$^3$, but the leakage plugging measure didn't efficiently function. After the density was further reduced to 1.87 g/cm$^3$, the leakage plugging measure successfully achieved the regular drilling until the well drilling was finished. The whole well circulation loss amount of drilling fluid was approximately 240 m$^3$, the work-time loss was 230 hours. After the operation of reduced density (1.87 g/cm$^3$), there was almost no more circulation loss fault. Figure 3 shows the density changing curves of W**2-8 well.

The designed original third spudding density of the W**2-7 well was 2.03 g/cm$^3$. Combined with the w**33-4 well and W**2-8 well operation experiences, the initial drilling fluid density of the third spudding directly was approved as 1.86 g/cm$^3$. The pressure-controlled drilling operation ran smoothly, and the drilling procedure was regular till the drilling was finished. Due to the initial density adopted the optimized density, there was almost no circulation loss failure in the third spudding during the W**2-7 well drilling operation.

Table 3 shows a time-sensitive comparison of horizontal segment drilling before and after the density reduction of the W**2 platform wells. It is shown that during the drilling in horizontal segment, the ROP of W**2-8 well with reduced density got the maximum value, and W**2-7 well drilling ROP came to the second; the drilling of W**2-8 well before reduced density implementation got the minimum ROP value, W**2-10 well drilling ROP was the second lower one. To sum up, the lower density of the two wells got the higher ROP data.

The ROP of density reduction measures well was around 20% higher than wells without density reduction measures on the same platform. This field test indicated that in the safety density window range, by reasonably reducing density redundancy, the lower density fluid drilling in southern Sichuan region was conducive to reducing the complexity of down-hole accidents, improving the ROP and reducing the general cost.

**Table 3** Time-sensitive comparison before and after density reduction of the W**2 platform wells

| The well number | Drilling fluid density g/cm$^3$ | Start depth m | End depth m | Horizontal depth m | Pure-drilling time hr | ROP m/hr |
|---|---|---|---|---|---|---|
| W**2-5 | 1.98-2.04 | 3178 | 4835 | 1657 | 314.32 | 5.27 |
| W**2-10 | 1.98-2.03 | 4237 | 5330 | 1093 | 110.5 | 8.45 |
| W**2-9 | 1.98-2.06 | 3397 | 4910 | 1513 | 177.5 | 8.52 |
| W**2-8(before density reduction) | 1.98 | 3272 | 3484 | 212 | 25 | 7.68 |
| W**2-8(after density reduction) | 1.87 | 3484 | 4880 | 1396 | 116.3 | 11.32 |
| W**2-7 | 1.86 | 3279 | 4705 | 1426 | 124.5 | 10.49 |

## 5. Conclusions and recommendations

The field experience of horizontal segment drilling operation of the hard-brittle shale in the Longmaxi formation and the field tests of HDOF shows that the reliability of the fluid performance will directly affect the stability of the wellbore wall. Regarding the drilling fluid, it can be concluded that:

(1) The countermeasures to maintain a higher osmotic pressure at the bottom-hole in the high-density drilling operation would benefit to reduce the formation pore pressure of the bottom hole, widen the fluid density window, and maintain the stability of the wellbore wall.

(2) By improving the formulation of HDOF, using new agents such as polymer modified nano-sealing agent, ultra-micro barite weight additive and optimizing the solid particle size distribution of drilling fluid can effectively improve the wellbore stabilization performance of HDOF and reduce the sloughing blocks of wellbore walls.

(3) The field test of "pressure-controlled drilling + density reduction" in horizontal drilling operations in southern Sichuan area is conducive to reducing the complexity of down-hole faults, improving the ROP, and saving the general cost.

(4) The density design should fully consider the adverse effects of the corresponding HDOF down-hole complexity circumstance and appropriately reduce the density redundancy, to definitely ensure that the key index work within the scope of safety density window as well.

In view of the above summarization and understanding of the possible problems in the shale gas extraction process of Weiyuan block, It should be paid attentions to the measures of strengthening the wellbore wall sealing，stabilizing the wellbore, controlling the solid sag, optimizing the high temperature performance of HDOF in horizontal segment operation, and taking some parallel measures to improve the quality of drilling service.

## Acknowledgements

Project supported by the National Major Scientific and Technological Special Project "Deep Well Ultra Deep Well High-Quality Drilling Fluid and Cementing Completion Technology Research" (No. 2016ZX05020-004).

## Conflict of interests

The author declares that there is no conflict of interest.